\documentclass[conference]{IEEEtran}

\usepackage{amsmath}
\usepackage{accents}
\usepackage{caption}
\usepackage{eqparbox}
\usepackage[pdftex]{graphicx}
\usepackage{subcaption}
\usepackage{physics}
\usepackage{lipsum}
\usepackage{svg}
\usepackage{xspace} 

\def \CPU {\textsc{cpu}}

\def \linkIdx      {\ell}
\def \linkSet      {L}

\def \linkCAP      {\textsc{link\_cap}}

\def \KPIvar    {\kappa}
\def \KPIvector {k}

\def \lpkpi     {LP-KPI\space}

\def \pMetric   {\mathcal{P}}

\def \RAM  {\textsc{ram}}
\def \rsrc {\textsc{rsrc}}

\def \StateIdx      {s}
\def \StateMatrix   {S}
\def \STO           {\textsc{sto}}

\def \Throughput   {R} 
\def \timeIdx      {t}
\def \timeWindow   {T}
\def \trafficmodel {LSTM-FSD\space}

\def \vnfIdx    {\nu}
\def \vnfSet    {\mathcal{V}}
\def \VNF       {\textsc{vnf}}

\begin{document}
\title{ML KPI Prediction in 5G and B5G Networks}

\author{\IEEEauthorblockN{Nguyen Phuc Tran}
\IEEEauthorblockA{Concordia University\\
Montréal, Québec, Canada \\
phuc.tran@mail.concordia.ca}
\and
\IEEEauthorblockN{Oscar Delgado}
\IEEEauthorblockA{Concordia University\\
Montréal, Québec, Canada \\
oscar.delgadocollao@mail.concordia.ca}
\and
\IEEEauthorblockN{Brigitte Jaumard}
\IEEEauthorblockA{Concordia University\\
Montréal, Québec, Canada\\
brigitte.jaumard@concordia.ca}
\and
\IEEEauthorblockN{Fadi Bishay}
\IEEEauthorblockA{Ciena\\
 Ottawa, ON, Canada\\
fbishay@ciena.ca}
}

\maketitle

\begin{abstract}
Network operators are facing new challenges when meeting the needs of their customers. The challenges arise due to the rise of new services, such as HD video streaming, IoT, autonomous driving, etc., and the exponential growth of network traffic.
In this context, 5G and B5G networks have been evolving to accommodate a wide range of applications and use cases. Additionally, this evolution brings new features, like the ability to create multiple end-to-end isolated virtual networks using network slicing.
Nevertheless, to ensure the quality of service, operators must maintain and optimize their networks in accordance with the key performance indicators (KPIs) and the slice service-level agreements (SLAs).

In this paper, we introduce a machine learning (ML) model used to estimate throughput in 5G and B5G networks with end-to-end (E2E) network slices.
Then, we combine the predicted throughput with the current network state to derive an estimate of other network KPIs, which can be used to further improve service assurance.
To assess the efficiency of our solution, a performance metric was proposed. Numerical evaluations demonstrate that our KPI prediction model outperforms those derived from other methods with the same or nearly the same computational time.

\end{abstract}


\IEEEpeerreviewmaketitle


\section{Introduction}

Research into the fifth-generation (5G) mobile networks and beyond (B5G) has captured the attention of both academia and industry. 
In this context, network slicing is an innovative new technology to provide personalized services and to meet the quality of service (QoS) requirements of individual customers. 
Thus, service assurance (SA) and network KPI monitoring are critical components of network slicing provisioning.
As 5G networks are virtualized to support next-generation applications, human troubleshooting cannot keep up with the new services and their increasing complexity.
Additionally, service providers require network service assurance methods to maximize QoS as well as to improve customer quality of experience (QoE).

Network slicing is a unique feature of 5G networks, that promises a significant quality of service improvement. It helps network operators to build logical (virtual) networks that can isolate resources and therefore improve the overall performance of the network \cite{8039298}.
Network slicing is defined as an end-to-end logical network with a group of isolated virtual resources on a shared physical infrastructure.
Thus, network function virtualization (NFV) and software-defined networking (SDN) are important concepts for developing 5G and B5G network slicing \cite{nig2018}.
SDN and NFV are used to dynamically create and manage network slices in order to provide customers with efficient, programmable, and scalable network services. 
Although the network slicing technology can enhance network performance and efficiency, it also brings new challenges: the deployment of network slicing, as well as managing the scaling, are some of the key difficulties \cite{son2019}. 
Network slice orchestration and deployment in a 5G network, therefore, constitutes a virtual network embedding problem \cite{tae2019}, in which virtual network components are mapped to physical nodes and virtual links are assigned to physical ones. 
In other words, we need to monitor KPIs to properly define and implement a network, as well as to guarantee the efficient functioning of the deployed network infrastructure. 
To fulfill the SLA contract, it is necessary for a network slice
to be adaptable to network traffic fluctuations, as well as to be able to minimize KPI violations while optimizing network resources to maximize profit.
Due to the significance of KPI measurements for service assurance, this paper will describe and implement a machine learning-assisted approach that estimates network KPIs based on network traffic.

Our contributions can be summarized as follows: \vspace{-0.2cm}
\begin{itemize}
    \item We propose a traffic forecasting model, {\trafficmodel}, at the network slice level based on network throughput and resource utilization.\vspace{-0.1cm}
    \item We introduce a method called \lpkpi to integrate the predicted throughput and the current network state information for predicting other KPIs.\vspace{-0.1cm}
    \item We define and implement a performance metric to evaluate the performance of \lpkpi.\vspace{-0.1cm}
    \item We evaluate and compare our proposed model to other machine learning algorithms.\vspace{-0.1cm}
\end{itemize}

The remaining sections of the paper are structured as follows:
section \ref{section-literature-review} covers the most recent methodologies used for traffic prediction and KPI forecasting; section \ref{section-system_design} discusses our machine learning-based method for traffic prediction, the mechanism for predicting additional KPIs, and the definition of our performance metric;
section \ref{section-experiment} demonstrates our experiments and comparisons with other methods;
finally, we describe our main findings and future work in section \ref{section-conclusion}.


\section{Literature Review}
\label{section-literature-review}


\subsection{5G Traffic Prediction}

As reflected in the mobility report \cite{Mobile}, the 5G network's proportion of mobile data traffic was around 10\% in 2021 and is projected to reach 60\% by 2027.
The mobility report provides valuable insights. It highlights the impact of network traffic patterns on the network infrastructure. 
Consequently, several recent studies have focused on traffic forecasting, which may help to better allocate network resources and to design better service assurance mechanisms. 
Traffic prediction could help to capture the dynamic and complicated behaviour of network traffic patterns, thus designing a comprehensive methodology for reliable traffic forecasting and enabling the combination of other components that can be used for network provisioning are important areas of research.
This is essential for network operators since it allows them to optimize network resources and to better dimension the network.
Le \textit{et al.} \cite{Le2018} demonstrated the relevance of network traffic and its connection with other KPIs; the authors also described in detail the significance of traffic forecast in mobile networks. They tested their proposed model on different networks (GSM, 3G, 4G) for both long-term and short-term forecasting.

In another study \cite{shr2021}, the authors evaluated the possibility of using machine learning on real-time 5G data and applied several ML algorithms on a data set derived from LTE network base stations. 
However, those experiments remained limited due to their reliance on the LTE data set, whose information may differ from the 5G network environment. 
In addition, the study did not take into consideration the slicing concept introduced in 5G and B5G networks.
Gao \cite{gao2022} implemented a deep learning neural network traffic prediction model with high accuracy, based on Long Short-Term Memory (LSTM). Their data set was generated in a 5G network environment with two modes: static and in-vehicle, and it mainly relied on a large bandwidth assumption. 
Even though the concept of network slicing was not mentioned in their research, it demonstrated the potential of traffic prediction based on neural networks, particularly LSTM.
In addition, LSTM models are effective in traffic forecasting due to their ability to capture long-term dependencies and handle time-series data, as demonstrated in studies such as \cite{wang2020long}.

Traffic prediction has demonstrated its applicability in networking, although its expansion to network slicing and how it can help to estimate other network KPIs such as delay, packet loss, jitter, etc. is still under investigation.
It can also be used to develop mechanisms to mitigate SLA violations on 5G and B5G networks.

\subsection{Machine learning techniques and service assurance}

It is anticipated that there will be more sophisticated SA requirements, such as the inclusion of specialized SA services provided by approved customers or third parties. 
The principal objective of a SA solution is to continuously monitor, measure, and evaluate real-time network traffic in order to maximize end-user QoE as defined by the SLAs. 
To handle the collected KPI data, more advanced analytic methods, such as Artificial Intelligence (AI), have been deployed to enhance SA in mobile networks \cite{xie2020}. 
To implement an effective SA solution in a 5G network, the operator must consider both slice orchestration and data management to guarantee SA in their network.
The authors in \cite{9282698}, introduced an orchestration mechanism to enable cross-slice communication to reduce infrastructure complexity and effectively utilize shared network service resources.

In other words, AI technologies, including ML, have shown enormous success in a wide range of application domains, as well as their potential to tackle the 5G network challenges \cite{tan2018}. 
In the world of big data today, there is a strong interest in this form of AI and the 5G/B5G network is not an exception; enterprises are required to deal with the enormous amount of information that their systems are now continually creating.
In the past, if there was a problem with the network, technicians would often perform drive tests and rely on their expertise and understanding of the industry to determine where issues occurred.
Additionally, the operation costs incurred by network operators were high and required a team to monitor and support 24/7, thus fixing the network problems can take a long time.
Consequently, we can say that AI/ML technology is one of the most powerful methods to proactively mine data and take network automation-related decisions.

Addressing the issue of how to use ML techniques to enhance network operation, the authors of \cite{9828381} implemented a 5G network environment based on an open-source MANO framework capable of managing and orchestrating Virtual Network Functions (VNFs).
Then, they investigated the network characteristics and presented an ML model, based on LSTMs, that was able to predict resource usage (CPU and RAM) of VNF instances in a network slice; allowing them to devise a threshold-based algorithm than defines the VNF resource consumption, but their proposed solution was not fully automated and had not touched on network KPIs as well as the link capacity of the E2E network slice.
Survey \cite{ma2020} delivered an overview of the state-of-the-art of online-data analytic tools that support proactive network optimization in 5G using ML, as well as, relevant information for optimization mapping and a description of the models used to forecast utilization patterns and correlate them with network KPIs.
In recap, we should notice that VNF resource usage is dependent on several other network variables and not only on itself \cite{8486246}.
In addition, we should remember that KPIs are the necessary metrics used to measure network performance and to take decisions, such as whether to increase or decrease the size of a network segment (VNF resource, transport link capacity, RAN resource, etc.).
In this paper, we address the research question of whether KPIs can be estimated based on network traffic demand.
Since access to a physical testbed is not easy to obtain and quite expensive, our analyses are based on a packet-level 5G simulation.


\section{Machine learning-assisted system design}
\label{section-system_design}

This section describes our proposed machine learning-assisted method for estimating network KPIs using our traffic prediction model.
First, we discuss how we construct a model for traffic forecasting, \trafficmodel \hspace{-1mm}, in a 5G E2E network environment.
Then, the methodology, \lpkpi \hspace{-1mm}, for estimating other KPIs is explained.

\subsection{Throughput KPI prediction}
\label{section-traffic-prediction}

In a 5G/B5G network with slicing enabled, each resource usage is not fully independent as it depends on the network status, such as current resource consumption, link capacity load, KPI values, etc., as well as configuration parameters such as maximum resource (CPU, RAM, Storage), guaranteed link rate, routing, etc.
Additionally, measuring network traffic and other KPIs are fundamental aspects of monitoring the performance of the network. 
Depending on the characteristics of the slice, a KPI can have distinct priorities; for example, a video streaming service may set a higher latency priority.
In addition, it takes time for each VNF that is operating in the network to become deployed and configured. 
Based on these insights, we propose a machine learning-based method for predicting short-term network traffic in 5G/B5G with E2E network slicing that takes network configuration and VNF resource utilization into account.
Traffic forecasting is essential since it mostly relies on user/application requests.

\begin{figure}[ht]
    \centering    
    \includegraphics[width=.9\columnwidth]{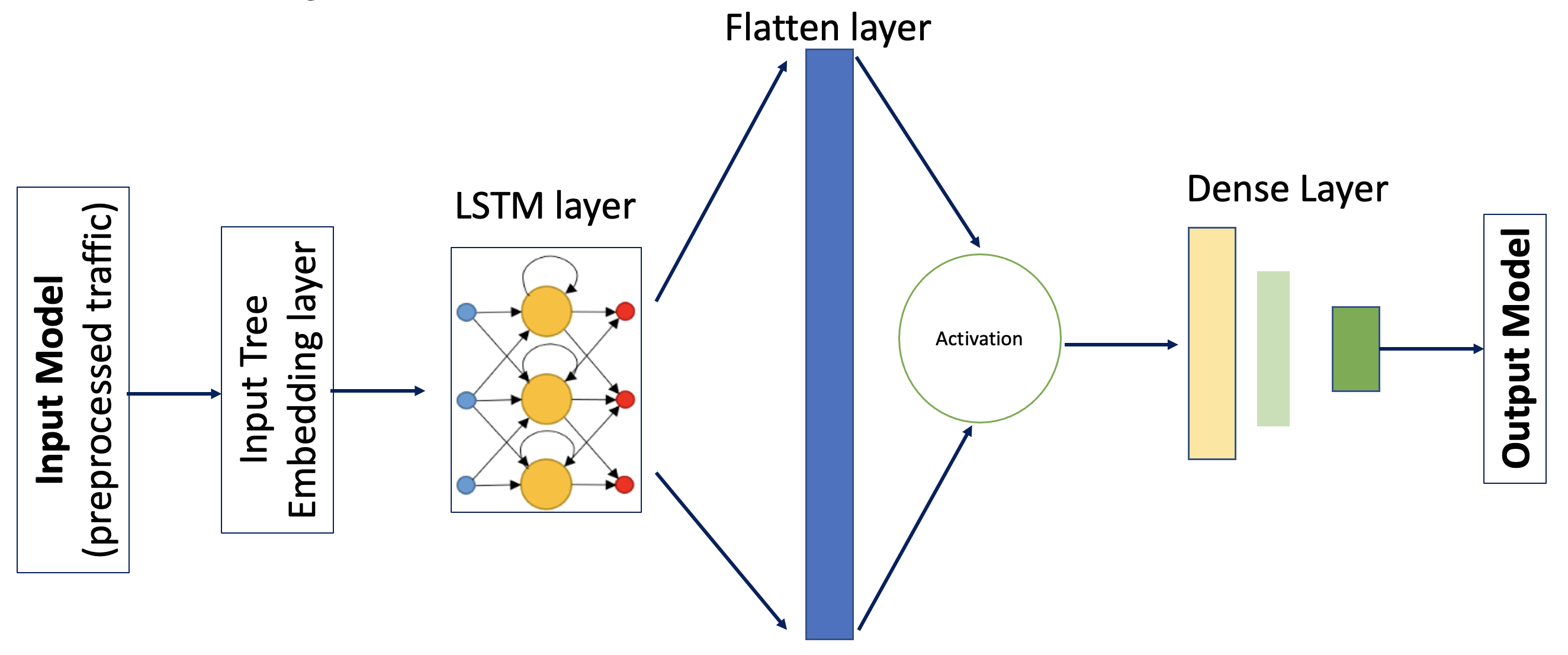}
    \caption{Traffic forecasting model (\trafficmodel).}
    \label{fig:rnn_lstm_model}
\end{figure}

Our traffic forecasting model \trafficmodel is constructed using an LSTM neural network since it's proven effective in traffic prediction.
The main features are extracted from the traffic history, the network state information and the slice configuration.
Detailed in Fig. \ref{fig:rnn_lstm_model}, our proposed prediction model aims to provide short-term traffic forecasting.
In order to increase precision and deal with the short-term data, we have used an LSTM layer combined with a Flatten layer and a Swish activation function, followed by a Dense layer as the output of the model.
For more details, Staudemeyer and Morris \cite{staudemeyer2019understanding} reported a comprehensive review of LSTM neural networks and their complexity analysis.
In the LSTM layer, a longer time window would require more resources for training and production; therefore, in our experiments, we selected a time window of 5 minutes to reduce the prediction response time while still meeting our goals.
In addition, the Flatten layer can help to deal with multi-dimensional inputs and to transfer the information through the activation function.
The Swish activation function is a smooth, non-monotonic function that always matches or beats ReLU on deep networks and has been used successfully in multiple difficult domains \cite{tri19}.
These techniques help our model avoid over-fitting and increase precision while still using short-term data.

\begin{figure}[ht]
    \centering   
    \includegraphics[width=.9\columnwidth]{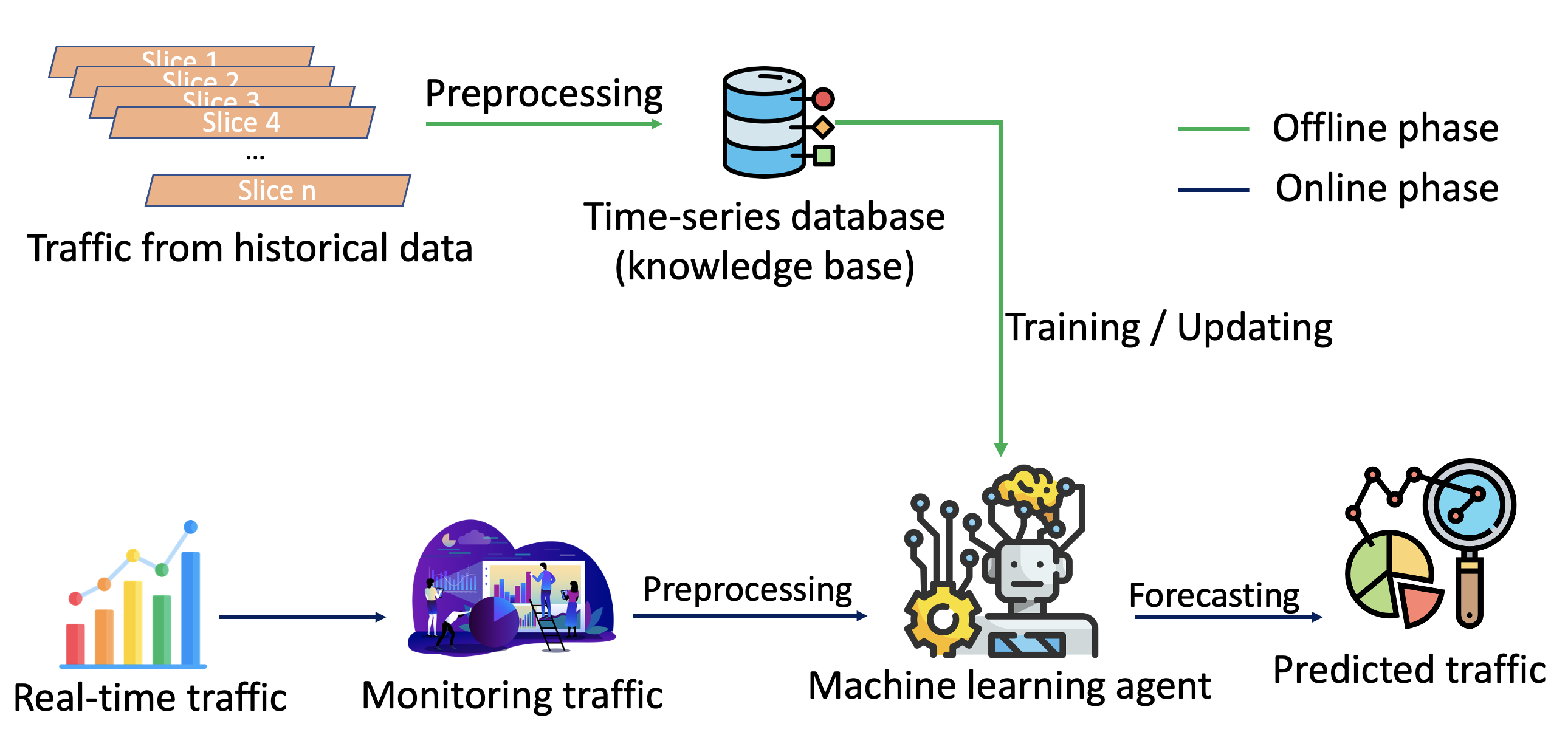}
    \caption{ML-assisted infrastructure workflow.}\vspace{-0.75cm}
    \label{fig:workflow_traffic_forecast}
\end{figure}

Depending on the available data set, a different time window value ($T$) can be used for building the time-series data.
Similar to other time-series models, one training observation at time $t$ is composed of 
$slice\_type$, 
$\Throughput$ (throughput), 
current VNF resource load ($\VNF^{\rsrc} = \{\CPU, \RAM, \STO\}$), and current link capacity load ($\linkCAP$) data collected within the time window; the output is the predicted $\widehat{\Throughput}$ at time $t+1$. 
The data used for training and evaluation can be easily captured at the radio access, transport or core network levels by network monitoring components. 
Fig. \ref{fig:workflow_traffic_forecast} depicts our comprehensive workflow, including traffic collection, training, and forecasting.
First, all of the information at each network slice, including throughput, configuration information, and VNF resource utilization, will be collected at monitoring components. 
Later, the collected data is pre-processed and stored in a time-series database as a knowledge base. 
Then, the data retrieved from the time series database is used to train and evaluate our proposed model. 
To obtain the predicted throughput once real-time traffic has been captured, the same pre-processing mechanism will be applied and fed to the machine learning model.



\subsection{Other KPI predictions}
\label{derive-additional-kpis}

Multiple KPIs, including throughput, delay, packet loss, jitter, stability and others, can be used to evaluate the 5G/B5G network performance and invoke a decision on the network operation. 
The relationship between VNF resources, traffic, and network KPIs was intensively investigated in \cite{9155263}.
VNF resources, such as CPU, memory, and storage, are used to process and manage the traffic that flows through the network.
If a VNF does not have enough resources, it may become overwhelmed and unable to process the traffic efficiently, leading to increased delay and packet loss. This can negatively impact the overall stability of the network.
As indicated earlier, the throughput prediction could be estimated by \trafficmodel \hspace{-1mm} but the reliability of the prediction has extreme significance; for instance, in the case of the delay, providing a forecast smaller than the actual value might cause the network to reduce its resources and consequently create a bottleneck.
In addition, with the assistance of predicted throughput, the management component can be combined with the current network information to estimate some KPIs of the network, and to dimension the network resources (e.g. link capacity, VNF resources, switch traffic to the edge network, etc), of each slice, ensuring the SLA requirements of each slice are met (delay, guarantee bit-rate, packet loss, etc.).
In order to accomplish a highly accurate prediction while leveraging ML models, it is essential to deal with large volumes of information. However, when using too complex or too many ML models, the procedure for deploying the solution changes inside the system, requiring additional effort and increasing the complexity of the network infrastructure.

To address these problems, especially for sensitive KPIs (e.g. delay, packet loss, jitter), we propose the \lpkpi model which minimizes underestimation and achieves good accuracy using the predicted throughput estimated by our ML agent described above.

Using a simplified ML model combined with our \lpkpi, rather than using a more complex ML model can improve performance and reduce computational costs.
Additionally, it optimizes the infrastructure and makes it simpler to interpret while still accomplishing the same accuracy as a complex ML model (or multiple ML models).
Our proposal seeks to predict other KPIs by utilizing the short-term predicted throughput and the current network state to assist network administrators in evaluating and adjusting the network configuration.
The idea of predicting additional KPIs can be described as follows:
\begin{equation}
\label{eq:min}
    \hat{\KPIvar} = \min \{ \StateIdx_{\timeIdx} \cdot w + \varepsilon : 
                            \StateMatrix w \geq \KPIvector \text{ for } \timeIdx \in \timeWindow, 
                            w \in {\rm I\!R} \}, 
\end{equation}
where:
\begin{itemize}
\item $\hat{\KPIvar}$: KPI to be estimated for the next time slot (e.g., delay, packet loss, jitter, etc),
\item $T$ is a time window ending at the current time slot,
\item $\StateIdx$: vector $[V^{\rsrc}, \widehat{\Throughput}]$ of network's state information and predicted throughput $\widehat{\Throughput}$ at next time slot. where $V_{\timeIdx}^{\rsrc} = (\VNF_{\timeIdx}^{\rsrc}, \linkCAP_{\timeIdx})$ is the VNF resource vector,
i.e., $\VNF_{\timeIdx}^{\rsrc} = (\VNF_{\vnfIdx, \timeIdx}^{\rsrc})_{\vnfIdx \in \vnfSet}$ is the compute resource vector at time $\timeIdx$ of all the VNFs $\vnfIdx$ in a given slice, and the last term is the link capacity vector at time $\timeIdx$ of all links $\linkIdx$ ($\linkCAP_{\timeIdx} = (\textsc{cap}_{\linkIdx, \timeIdx})_{\linkIdx \in \linkSet}$),
\item $\StateMatrix$: matrix of network's state information and throughput $\Throughput$ for the time window period. 
Each row corresponds to vector $[V^{\rsrc}_{\timeIdx},\Throughput_{\timeIdx}]$ for $\timeIdx \in \timeWindow$, 
\item $w$: weight vector,
\item $\KPIvector$: vector of the previous values of the KPI to be estimated, for $\timeIdx \in \timeWindow$,
\item $\varepsilon$: noise parameter.
\end{itemize}

To make the derived KPI reliable and to guarantee there is no underestimation, which could result in SLA violations, the constraint in equation \ref{eq:min} combines the current network state information and the historical states.
Then, the estimated KPI value is obtained by solving the optimization problem utilizing the predicted throughput value obtained by \trafficmodel \hspace{-1mm}.
One advantage of our proposed model is that it can be readily integrated with other components, such as SA algorithms, network planning, assist/validate other ML models, etc.


\subsection{Performance metric}
\label{metric}

Below we define a performance metric $\rho$ to assess the accuracy of \lpkpi model prediction against various classic ML algorithms.
The idea of $\rho$ is to provide a way to measure the accuracy of our estimations by taking into consideration over and under-estimations.
\begin{equation} 
\label{eq:error_metric}
    \rho = 
    \begin{cases}
        1,      & \text{if $d = \frac{\hat{\KPIvar}-\KPIvar}{\hat{\KPIvar}+\KPIvar} \in ( 0, \bar{d} ]$ } \\
        3,      & \text{if $d > \bar{d}$} \\
        5,      & \text{if $d < 0$} \\
        0,      & \text{otherwise,}
    \end{cases}
\end{equation}
where 
$\bar{d}$ is the acceptable maximum over-provisioning limit for KPI $\KPIvar$, $\hat{\KPIvar}$ is the predicted KPI value, and $\KPIvar$ is the actual KPI value obtained at the same timestamp.

At a given time, a normalized gap value $d$ between the derived value $\hat{\KPIvar}$ and the actual value $\KPIvar$ will be calculated.
We also need to define an acceptable value of the over-provisioning limit $\bar{d}$, e.g., $\bar{d}=0.15$ (equal to 15\%).
Since inaccurately estimating a KPI might create problems when dimensioning the network, our performance metric assigns a penalty of 5 points when the derived KPI is underestimated.
On the other hand, when a KPI is estimated within an acceptable range $\bar{d}$, such as when network resources are properly configured, then the network is able to serve in a normal state; thus, our metric gives a small penalty of 1 point.
In addition, if the KPI is overestimated, 3 points are assigned. 
Finally, the penalty for high accuracy (equal or nearly equal to the actual value) is 0.
Depending on the data set, the number of observed timestamps $n$ could be several hours, a day, or the whole week.

The overall performance $\pMetric$ can be estimated as follows:
\begin{equation} \label{eq:error_rate}
    \pMetric = \frac{1}{n} \sum\limits_{\timeIdx \in \mathcal{T}} {\rho}_{\timeIdx} 
\end{equation}
where $n = \vert \mathcal{T} \vert$ is the number of time slot events and 
$\rho_{\timeIdx}$ is the performance metric at time $\timeIdx$.

We have selected 5 ML algorithms which are able to predict values in a reasonable amount of time as a benchmark to evaluate our \lpkpi model. 
This is important in a 5G/B5G network because any model needs to predict the value reasonably quickly for the network operator to have sufficient time to adapt the network and compute resources. 
The details of our experiments are described in Section \ref{section-experiment}.
\section{Experiments}
\label{section-experiment}

\subsection{5G environment and dataset}

\begin{figure}[ht]
    \centering
    \includegraphics[width=.9\columnwidth]{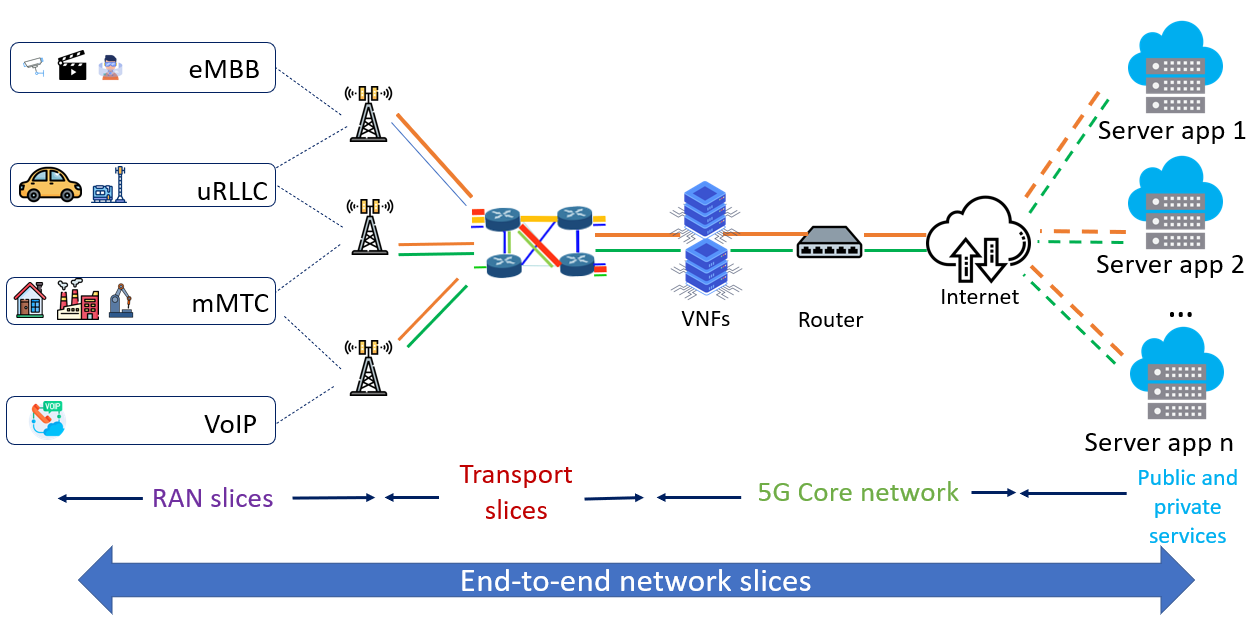}
    \caption{5G end-to-end network slicing.}\vspace{-0.5cm}
    \label{fig:5g_e2e_net}
\end{figure}

The 5G network described in this article was implemented on the 5G network level simulator OMNET++ (packet level).
The details of how to set up our 5G simulator and the implementation of the new features like network slicing can be found in \cite{del22}.
Fig. \ref{fig:5g_e2e_net} shows our high-level 5G network design including the radio access network (RAN), the transport and the core domains; the VNFs represent the user plane function (UPF) in our simulations.
The simulation is developed with four typical slice categories for a 5G network \cite{cha21}: Enhanced Mobile Broadband (eMBB), Ultra Reliable Low Latency Communication (uRLLC), Massive Machine-Type Communications (mMTC) and VoIP.

\begin{table}[ht]
    \centering
    \resizebox{0.8\columnwidth}{!}{
    \begin{tabular}{ | c || c | c| }
         \hline
         Slice   & Total UE & Service \\ [0.5ex] 
         \hline
         eMBB & 24,322 & HD video streaming. \\ 
         uRLLC & \phantom{1}7,304 & IoT applications. \\
         mMTC & \phantom{1}6,283 & Video game streaming. \\
         VoIP & \phantom{1}2,684 & VoIP services. \\
         \hline
        \end{tabular}
        }
    \caption{Number of user equipment for each slice.}\vspace{-0.5cm}
    \label{tab:map_vehicle_to_device}
\end{table}

To create a realistic 5G network scenario, a transportation data set from October 1, 2018, to October 8, 2018, from the City of Montreal \cite{opendatamtl} was used to simulate the traffic information.
Table \ref{tab:map_vehicle_to_device} shows the number of user equipment (UE) and application services defined in our 5G network environment.
Each UE in a slice is associated with a given application service that is configured for its slice's specific requirements.
For example, UEs are typically associated with high-bandwidth services such as streaming video and online gaming, while laptops are typically associated with business-oriented services such as cloud computing and video conferencing. 
By using network slicing, we can create dedicated slices of the network for each device type and application service, ensuring that the specific requirements of each device and service are met.
Due to the memory and computing limitations of the OMNET++ simulator, we have employed scaling factors based on the application type to reduce the number of devices in our simulation while retaining data integrity.
Finally, a time window of $T=5$ minutes was applied to our data set which was obtained from our 5G simulation in order to construct the time series described in the previous section. 


\subsection{Traffic forecast evaluation}

\begin{figure}[ht]
    \centering   \includegraphics[width=.8\columnwidth]{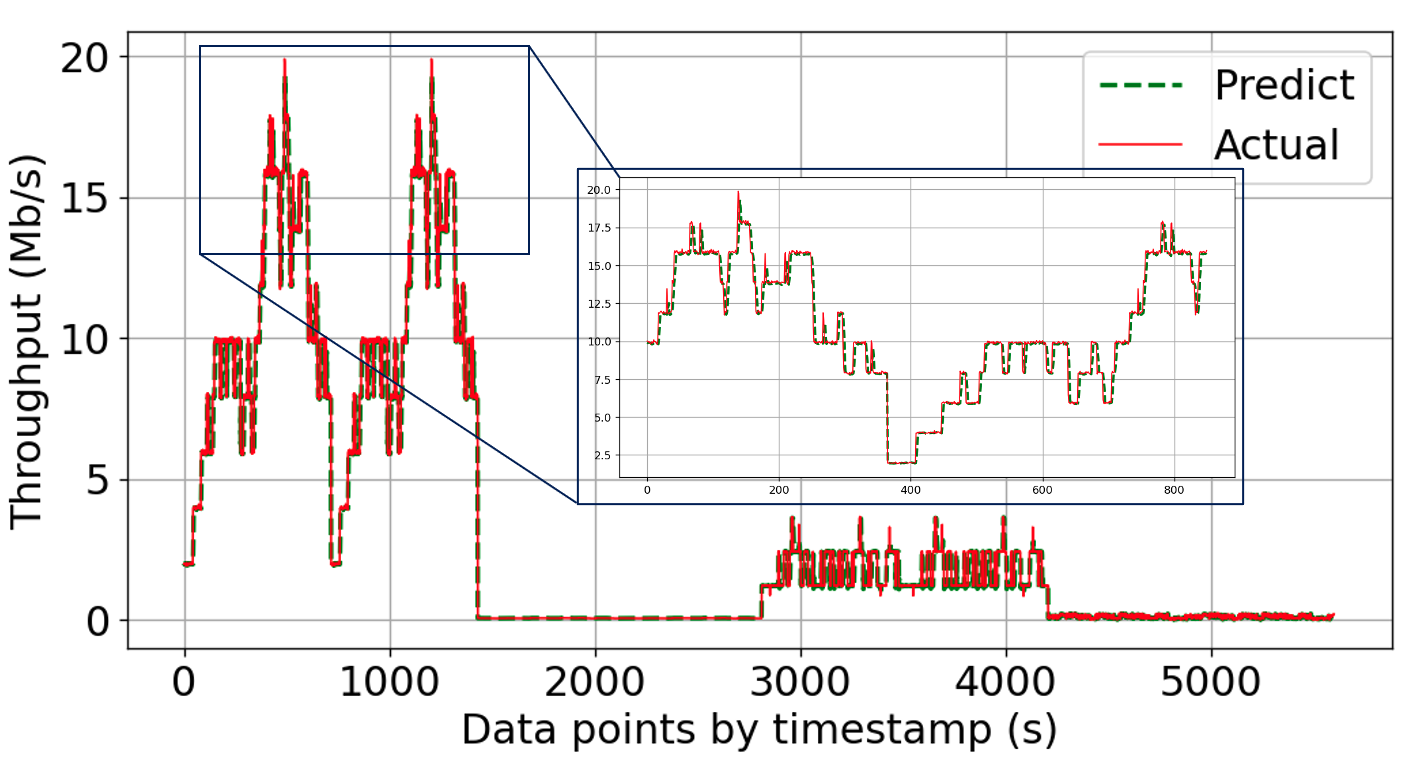}
    \caption{Data visualization when using \trafficmodel on traffic from all slices with T = 5 minutes.}\vspace{-0.7cm}
    \label{fig:test_model_lsm_plot}
\end{figure}

To evaluate the \trafficmodel model, we utilize data from different days and the mean absolute percentage error (MAPE). The MAPE formula can be defined by:
\begin{equation} \label{eq:mape}
    \text{MAPE} =\frac{1}{n}\sum_{\timeIdx \in \mathcal{T}} { \abs{ \frac{\KPIvar_{\timeIdx} - \hat{\KPIvar}_{\timeIdx}} {\KPIvar_{\timeIdx}} }  } 
\end{equation}
where $n = \vert \mathcal{T} \vert$ is number of time slot events, $\KPIvar_t$ is the KPI value $\KPIvar$ at time $t$ and $\hat{\KPIvar}$ is the forecast value of KPI $\KPIvar$ at time $t$. 
According to the training results, the traffic forecast model begins to converge around epoch 15 reaching a MAPE $=5.5\%$. 
When evaluated on a different day, \trafficmodel was able to perform well in terms of capturing the trend of the traffic (see Fig. \ref{fig:test_model_lsm_plot}), including some points with a high peak.
To compare our results, 5 ML algorithms were chosen that fitted our time series data: Linear regression, ARD regression, Ridge regression, Elastic net and Gaussian mixture, and we conduct and evaluate our experiments using the same data set.

\begin{table}[ht]
    \centering
    \resizebox{0.6\columnwidth}{!}{
    \begin{tabular}{ | c | l | c |  }
         \hline
         \# & Model & MAPE (\%) \\ [0.5ex] 
         \hline
         1 & \trafficmodel      & \textbf{18.66} \\ 
         2 & Linear Regression  & 52.62 \\
         3 & ARD regression     & 34.0 \\
         4 & Ridge regression   & 52.61 \\
         5 & Elastic net        & 67.29 \\
         6 & Gaussian mixture   & 67.29 \\[1ex] 
         \hline
    \end{tabular}
    }
    \caption{Traffic prediction model comparison.}\vspace{-0.5cm}
    \label{tab:traffic_prediction_model_comparison}
\end{table}

Table \ref{tab:traffic_prediction_model_comparison} provides a detailed overview of our findings.
In all cases, the algorithms were trained using one day of data and evaluated using an additional day of data.
Based on the results, we can demonstrate that our approach is feasible and can help to achieve high accuracy for short-term traffic prediction in 5G networks, with MAPE = 18.66\%. 
Due to the complexity of reallocating network resources, achieving a highly accurate traffic prediction is advantageous to network slicing-based applications and could avoid delays when scaling the network (either increase or decrease). Not only that but perhaps it is also a benefit when providing zero-touch proactive service assurance in the 5G/B5G network.


\subsection{\lpkpi model evaluation}

We now show the accuracy of the estimation of additional KPIs. The estimations are derived from the predicted traffic acquired from our ML agent \trafficmodel \hspace{-1mm}. 
We understand that evaluating different KPIs can vary depending on the metrics used and various network configurations. In this experiment, our initial focus is on assessing and investigating delay, packet loss, and jitter on four common slice types: eMBB, uRLLC, mMTC and VoIP.
Note that we evaluated and collected performance metrics using historical data collected on a day never used for training.
First, the collected data is used in the \lpkpi model, as well as a training set for other ML algorithms
The properties of this data set are described in Section \ref{derive-additional-kpis}. 
Then, the same data is utilized to feed our traffic prediction model and to obtain the predicted traffic $\widehat{\Throughput}$. 
Finally, the current network state is merged with $\widehat{\Throughput}$ to estimate the other KPIs and to compute the performance metric associated with each model.

\begin{figure*}[ht]
    \centering
    \begin{subfigure}{.7\columnwidth}
        \includegraphics[width=\columnwidth]{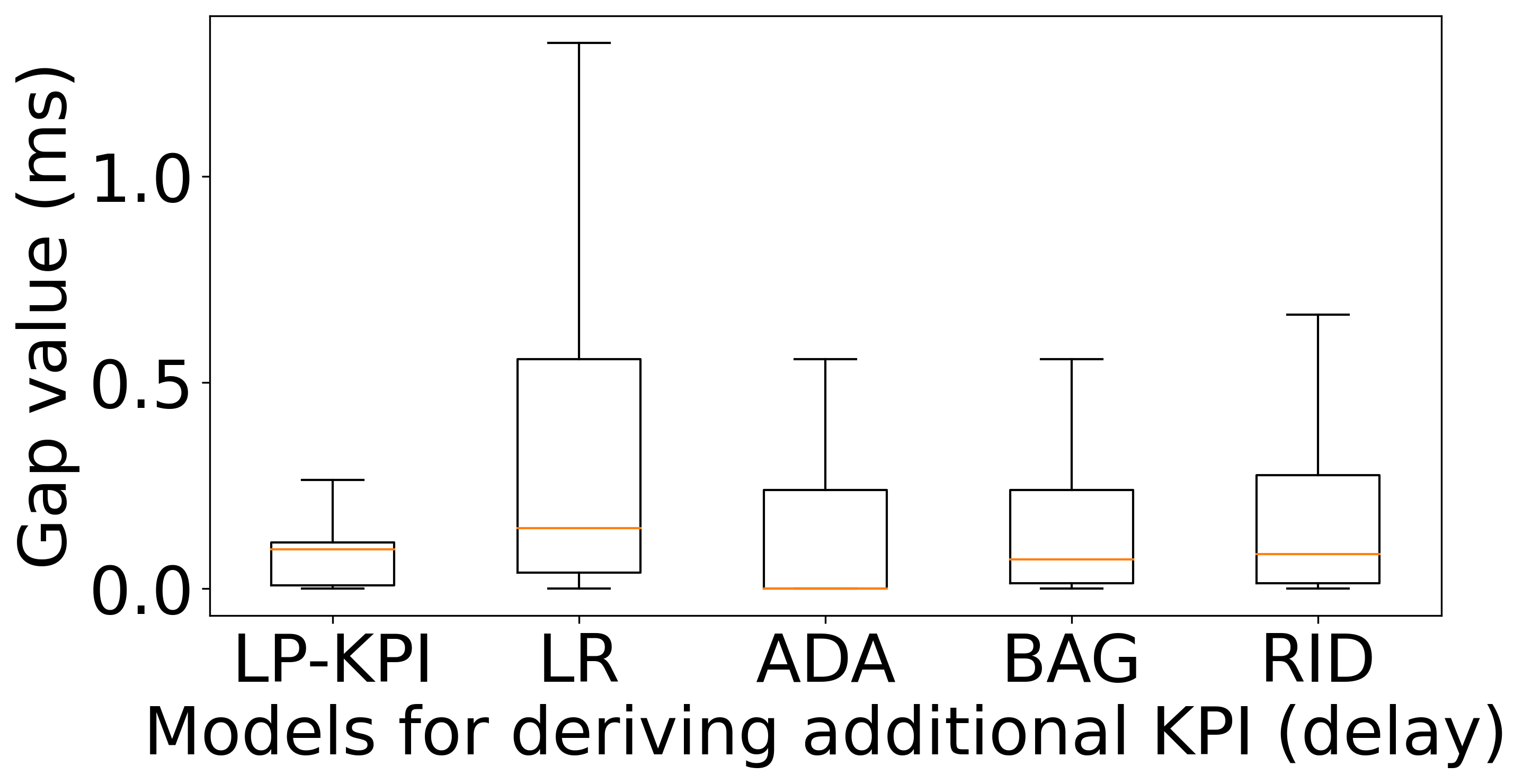}
        \caption{Delay.} \vspace{-0.5cm}
    \end{subfigure}
    \hspace*{1.cm}
    \begin{subfigure}{0.7\columnwidth}
        \includegraphics[width=\columnwidth]{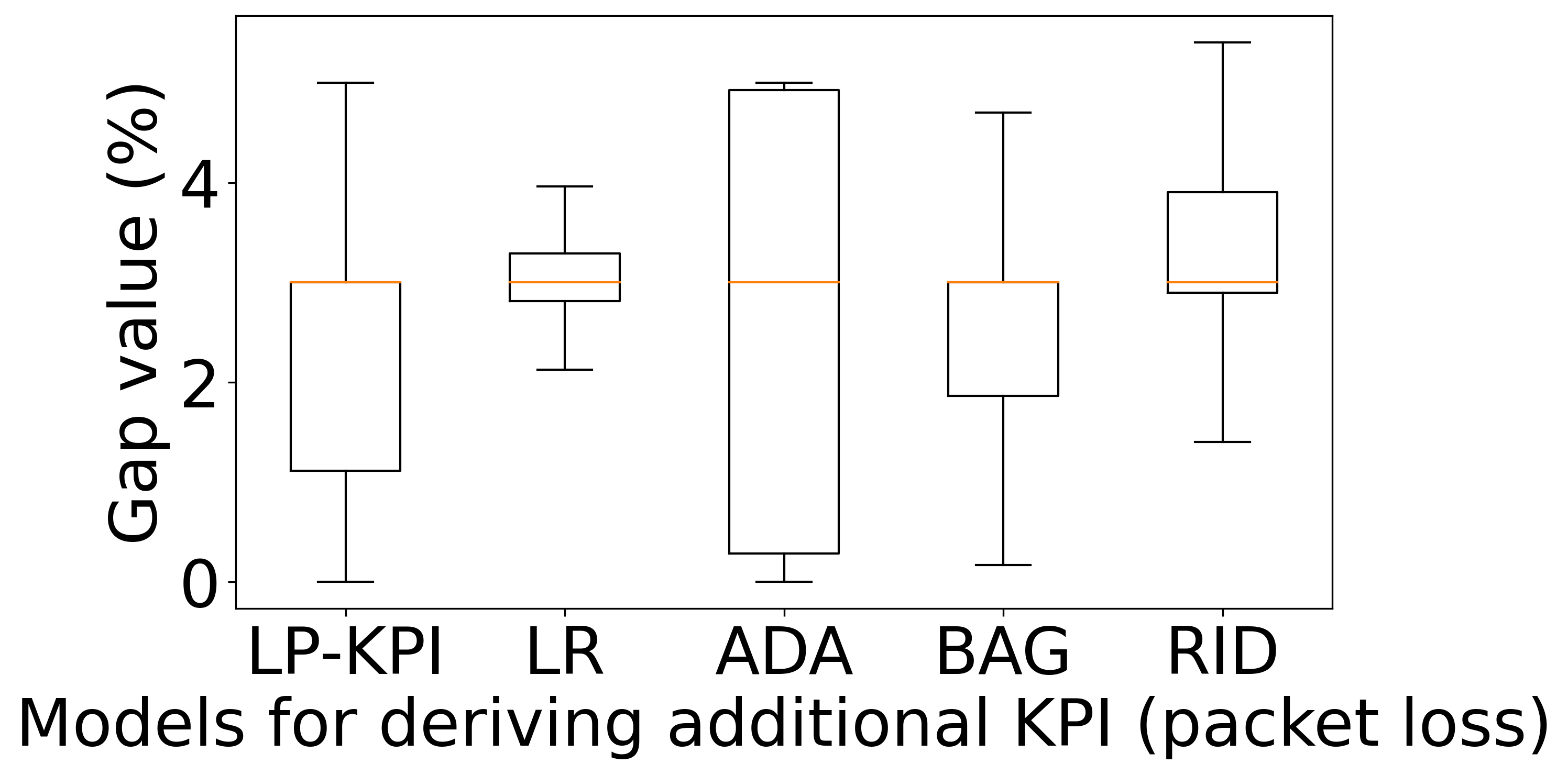}
        \caption{Packet loss.} \vspace{-0.5cm}
    \end{subfigure}
    \caption{Gap comparison for each model \protect (LR: linear regression, ADA: AdaBoost regressor, BAG: Bagging regressor, RIG: Ridge (Linear least squares with L2 regularization)} \vspace{-0.7cm}
    \label{fig:compare_gap_lp}
\end{figure*}

To evaluate all the models in our test environment, an acceptable amount of over-provisioning, $\bar{d}=15\%$, was defined. 
In our initial tests, we estimate the end-to-end latency and packet loss KPIs.
Fig. \ref{fig:compare_gap_lp} demonstrates the performance based on the  $\pMetric$; \lpkpi achieves high performance with minimal over-provisioning and under-provisioning on both delay and packet loss KPIs.
Our method has better accuracy and the smallest standard deviation from the actual KPI value (Gap).

Other aspects can be found in Table \ref{tab:derive_kpi_error_rate_comparison}. Our methodology has values of $\pMetric = 0.153$ for the delay KPI, and $\pMetric = 2.223$ for the packet loss KPI, which is more accurate than alternative ML techniques with higher error rates. 
\begin{table}[ht]
    \centering
    \resizebox{0.7\columnwidth}{!}{
    \begin{tabular}{ | c | l | c | c |  }
         \hline
         \# & Model & $\pMetric$-delay & $\pMetric$-packet loss \\ [0.5ex] 
         \hline
         1 & \lpkpi             & \textbf{0.153} & \textbf{2.223} \\ 
         2 & Linear regression  & 0.372 & 4.904 \\
         3 & AdaBoost regressor & 0.178 & 2.597 \\
         4 & Bagging regressor  & 0.169 & 2.663 \\
         5 & Ridge & 0.217 & 4.626 \\ [1ex]
         \hline
    \end{tabular}
    }
    \caption{Gap value comparison.} \vspace{-0.8cm}
    \label{tab:derive_kpi_error_rate_comparison}
\end{table}

\begin{figure}[ht]
    \centering  \includegraphics[width=.9\columnwidth]{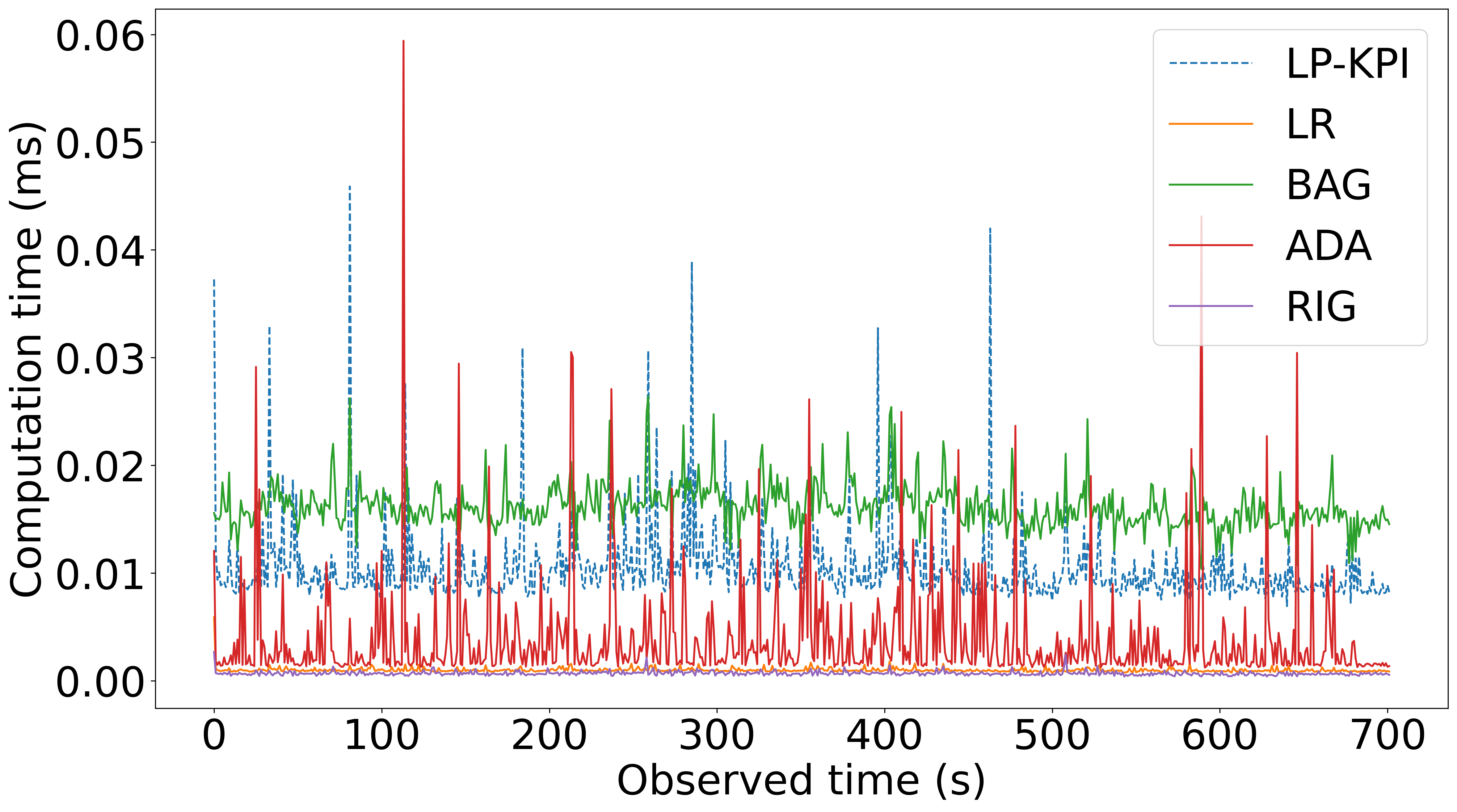}
    \caption{Computational times.} \vspace{-0.5cm}
    \label{fig:computation_time_derive_kpis}
\end{figure}

In addition, \lpkpi is a lightweight, simpler, and more easily implemented method. Thus, deployment cost is cheaper and faster compared to the other ML models. 
In terms of computational time, we found that \lpkpi was in the median (between RIG and BAG - see Fig. \ref{fig:computation_time_derive_kpis}), with an acceptable time as compared to RIG (the best ML method) of roughly 80~$\mu s$.


\section{Conclusion and future work}
\label{section-conclusion}

In this paper, we present and evaluate a novel traffic forecasting approach in 5G and B5G networks that accurately predicts short-term traffic throughput. 
We have also developed a new and reliable methodology for deriving other network KPIs estimations. Additionally, we have proposed a new performance metric to assess the quality of our estimations.
Our experiments demonstrate that our traffic forecasting model \trafficmodel, and the derived KPI estimation model \lpkpi, has good accuracy and low computational time.
Future directions of research include the use of our traffic forecasting model, in conjunction with the derived KPI model to design a zero-touch service assurance solution.

\section*{Acknowledgment}

First two authors were supported by a MITACS \& Ciena internship.


\bibliographystyle{IEEEtran}
\bibliography{IEEEabrv,Biblio/references,Biblio/Close_Loop_SErvice_Assurance,Biblio/Traffic_Prediction,Biblio/ML,Biblio/Online_learning, Biblio/E2E_NS}

\end{document}